\documentclass[runningheads,citeauthoryear]{apinv}
\usepackage{epsfig,cite,graphics}
\usepackage{marvosym} 
\usepackage[utf8]{inputenc}
\usepackage{color}
\usepackage[normalem]{ulem}

\begin{document}

\title{Geodesic model complience with the  \\
frequencies of the observed X-ray quasi-period oscillations of XTE J1807-294}
\titlerunning{How to prepare a paper for BlgAJ}
\author{Radostina Tasheva and Ivan Stefanov}
\authorrunning{R.Tasheva and I.Stefanov}
\tocauthor{Radostina Tasheva and Ivan Stefanov}
\institute{Department of Applied Physics, Technical University of Sofia, 8 St. Kliment Ohridski Blvd., BG-1000, Sofia \newline
	\email{rpt@tu-sofia.bg}    }
\papertype{Submitted on xx.xx.xxxx; Accepted on xx.xx.xxxx}	
\maketitle

\begin{abstract}
The investigation of the data for quasi-periodic pulsations observed in the X-ray spectra of the accreting millisecond pulsar XTEJ 1807-294 allows some conclusions to be made about its main parameters – mass and angular momentum.
Seven different geodesic models – namely RP, RP1, RP2, TP, TP1, WD and TD are applied in attempt to assess their ability to describe the properties of the central neutron star.
\end{abstract}

\section*{Introduction}
The field of X-ray timing was revolutionized by the launch of Rossi X-ray Timing Explorer (RXTE) on the 31st of December 1995, a space observatory carrying the Proportional Counter Array (PCA). The latter is an X-ray detector with a timing resolution down to milliseconds that allows extraction of X-ray data for stars in latest stages of their evolution.

Most of the information that we have about the observed low mass X-ray binaries (LMXB) is obtained through the studying of the X-ray spectra emitted by the accretion disk surrounding the central object, an  already evolved initially massive star – a black hole (BH) or a neutron star (NS). Neutron stars and black holes posses extremely strong gravitational field that can be tested through the exploration of the accreted gas motion and emission. The emission is in X-ray range, it is persistent and proves to be quasi-periodic. The observed frequencies can be naturally connected to the geodesic frequencies of motion of a test particle in the accretion disk. It is possible, of course that the quasi-periodic oscillations (QPOs) arise as a result of the intrinsic properties of the accretion disk flow rather than its geometry - for example magnetically-driven density waves in the disk, (Tagger\& Pellat, 1999). The geodesic models are simpler, though, and hence - more attractive. 

QPOs can be  devided in two groups – low-frequency (LF QPOs) and high-frequency (kHz QPOs). An advantage of LF QPOs, which tend  to drift in frequency, is their strength,  kHz QPOs are weak and rarely observed but have one significant advantage – some of them appear in pairs - a lower $\nu_{L }$  and a higher $\nu_{U }$ with rational frequency ratio (for example $3:2$ or $3:1$). Such a relation allows the equations that connect these frequencies to the basic properties of the central objects - NS or BH to be solved and a relation between the mass and the angular momentum of the object in question to be obtained.

This mass-angular momentum relation allow one of these parameters to be evaluated, if the other is known from additional data (for example, photometry). We are set to solve another problem – by testing different geodesic models to find the one that is the most suitable for the description of the frequenceis of the QPOs observed in the X-ray spectrum of XTE J1807-294 and, hopefully, for a certain class of objects.

Linares et al. (2005)  reported in 2005 for the first time twin kHz QPOs of the X-ray flux of the low mass binary XTE J1807-294. They observed eight different pairs of simultaneous kHz QPOs. In the present paper we apply seven different models that can be found in the literature – namely the relativistic precession model (RP), and its two modified versions (RP1) and (RP2), the total precession model (TP) and its modified version (TP1), the tidal disruption model (TD) and the warped disk model (WD),  in an effort to explain the values of the observed frequencies of the QPOs.  The optimal mass $M_{ opt}$ and optimal angular momentum $a_{ opt}$ of the object are obtained  with the help of the $\chi^{2}$ goodness-of-fit test. 

The paper is organized as follows. After presenting the data of the  twin kHz QPOs in XTE J1807-294 in Section 1 we proceed with a brief description of the models we intend to use in Section 2. The description of the $\chi^{2}$ test application and the discussion of the obtained results are presented in Sections 3 and 4 subsequently. The last section is the conclusion.

The masses received are  in solar masses, the radii $x$ are scaled with the gravitational radius i.e. $ x = {r}/{r_g}$  , where ${r_g}=GM/{c^2}$. The specific angular momentum  is  $a=c J/G{M^2}$ and, as usual for general relativity,  $G = 1 = c$ where $G$ is the universal gravitational constant and $c$ is the speed of light.

\section*{1. Observational data}

The first detection of  XTE J1807–294  was on 13th of  February, 2003,  during the periodic scans of the Galactic bulge region by means of  RXTE (Markwardt et al., 2003). The object was  in an outburst that lasted approximately 120 days. Linares et al. (2005) reported the presence of twin QPOs using data from 27th of February to 16th of March 2003. A multi-Lorentzian function is used to fit the power spectrum of each of the eight groups.
Table 1 enlists the eight groups of data and their corresponding frequencies  given in  (Linares et al., 2005). Seven of them seem to exhibit simultaneous QPOs. The first sample, A has to be treated with caution since the identity of the lower kHz QPO is not clear in set A.

\begin{table}[htb]
  \begin{center}
  \caption{Twin kHz QPO frequencies with their uncertainties }
\begin{tabular}{ c  l  l}
group                                            & $\nu_{L}$, [Hz]                                            & $\nu_{U}$, [Hz]  \\
\hline \\
A & $106.0\pm23.0$  & \,\,$337.0\pm10.0$ \\
B                                                   & $163.0\pm23.0$                                            & \,\,$354.0\pm4.0$  \\
C                                                   &  $191.0\pm9.0$                                               & \,\,$375.0\pm2.0$      \\
D                                                   &  $202.0\pm11.0$                                               & \,\,$395.0\pm3.0$      \\
E                                                   &  $238.0\pm25.0$                                               & \,\,$449.0\pm9.0$      \\
F                                                    &  $259.0\pm16.0$                                               & \,\,$465.0\pm2.0$      \\
G                                                   &  $273.0\pm19.0$                                               & \,\,$492.0\pm5.0$      \\
H                                                   &  $370.0\pm18.0$                                               & \,\,$565.0\pm5.0$      \\
                     \end{tabular}
  \label{twin}
  \end{center}
\end{table}
\section*{2. Geodesic models}
When massive stars, such as neutron stars or black holes, in binary systems reach their final stage of evolution an accretion
 disk is formed around them due to the matter in falling from the companion star. The accreted matter is moving  along slightly eccentric  orbits close to the innermost stable circular orbit (ISCO) and its emission could be responsible, on the first place, for the QPOs in the X-ray power-density spectrum. Therefore, the fundamental frequencies of motion of test particles flowing with the accreted matter, namely, the orbital frequency $\nu_{\phi}$, the radial $\nu_{r}$ and the vertical $\nu_{\theta}$ epicyclic frequencies, and simple combinations of them could be responsible for the frequencies exhibited in the X-ray spectrum. The role of the test particle is usually played by a conglomerate of hot blobs flowing with the stream of matter attracted  by the central object. 

In the “relativistic precession” model RP proposed in (Stella \& Vietri,~1998; Stella et al.,~1999; Merloni et al.,~1999) the upper frequency $\nu_{U }$ is considered a direct result of the modulation of the X-ray flux by the orbital frequency of the blob and $\nu_{U }$=$\nu_{\phi}$. The lower frequency $\nu_{L}$ coincides with the periastron precession of the relativistic orbit   i.e. $\nu_{per}$ = $\nu_{\phi} - \nu_{r} $. According to the RP1 and the RP2 models the precession mode and an oscillation mode of a hot inhomogeneity flowing in a slightly eccentric torus interact thus creating  a resonance responsible for the X-ray emission variability - (Bursa, 2005). According to the RP2 model, (Torok et al., 2011; Torok et al., 2012) kHz QPOs are produced due to resonance between the radial and the vertical modes.

In the tidal disruption model (TD)  the hot orbiting blobs are distorted by the tidal forces of the central object and form arches responsible for the observed modulation of the flux, (Cadez et al., 2008; Kostic et al., 2009; Germana et al., 2009).

According to “warped disk model” (WD) (Kato, 2004a; Kato, 2004b; Kato, 2004c; Kato, 2005a; Kato, 2005b) a resonance  due to a distortion in the accretion disk is proposed as an excitation mechanism for the orbiting particle oscillations.

The two total precession models TP and TP1 emphasize on the significance of the vertical precession mode which is causing the resonance responsible for the QPOs (Stuchlik et al., 2007). The total precession frequency is $\nu_{T }$ = $\nu_{per } - \nu_{LT} $=$\nu_{\theta} - \nu_{r}$, where $\nu_{LT}$ is the Lense-Thirring frequency or the nodal precession frequency of the orbital plane. The total precession $nu_{T }$  frequency corresponds to the time interval for which the declination of the quasi-eccentiric plane and position of the periastron reach the initial state simultaneously. The TP1 model utilizes $\nu_{U }$=$\nu_{\phi}$  instead  of $\nu_{U }$=$\nu_{\theta}$ for TP model.

The prescriptions of the used models for the frequencies of the kHz QPOs are summarized in Table 2.

\begin{table*}
	\centering
	\caption{Models for the kHz QPOs. References: 
		1. Stella \& Vietri,~1998;
		2. Stella et al.,~1999; 
		3. Merloni et al.,~1999;
		4. Bursa, 2005;
		5. Torok et al., 2011;
		6. Torok et al., 2012;
	    7. Stuchlik et al. 2007;
		8. Cadez et al., 2008;
		9. Kostic et al., 2009;
		10. Germana et al., 2009;
		11. Kato, 2004a;
		12. Kato, 2004b;
		13. Kato, 2004c;
		14. Kato, 2005a;
		15. Kato, 2005b.}
	\begin{tabular}{ |c|c|c|c|}
		\hline
		Model &$\nu_{L}$& $\nu_{ U}$& Ref. \\
		\hline
		\multicolumn{4}{|c|}{Relativistic precession models}\\
		\hline
		~~RP  & $\nu_{\phi}-\nu_{ r}$ & $\nu_{ \phi}$& 1, 2, 3\\
		~~RP1   & $\nu_{ \phi}-\nu_{ r}$ & $\nu_{ \theta}$ & 4\\
		~~RP2   & $\nu_{ \phi}-\nu_{ r}$ & $2\nu_{ \phi}-\nu_{ \theta}$  & 5, 6\\
		\hline
		\multicolumn{4}{|c|}{Total precession models}\\
		\hline
		~~TP  & $\nu_{ \theta}-\nu_{ r}$ & $\nu_{ \theta}$& 7\\
		~~TP1   & $\nu_{ \theta}-\nu_{ r}$ & $\nu_{ \phi}$ & 7\\
		\hline
		\multicolumn{4}{|c|}{Tidal disruption model}\\
		\hline
		~~TD   & $\nu_{ \phi}$ & $\nu_{ \phi}+\nu_{ r}$ & 8, 9, 10\\
		\hline
		\multicolumn{4}{|c|}{Warped disk model}\\
		\hline
		~~WD    & $2(\nu_{ \phi}-\nu_{ r})$ & $2\nu_{ \phi}-\nu_{ r}$ & 11, 12, 13, 14, 15\\
		\hline
	\end{tabular}
	\label{table_models}
\end{table*}

\section*{3. $\chi^{2}$ Goodness-of-fit test}
\subsection*{3.1 Recipe for the evaluation of the model function }
The   $\chi^{2}$ test is a powerful tool for consistency check. Different models yield different mass-angular momentum relations and subsequently different main parameters of the central object - mass $M$ and specific angular momentum $a$ . The  $\chi^{2}$ test allows us to distinguish between the models and to choose the best fitting one.

We have data that consist of eight pairs of values for the lower $\nu_{L }$   and the higher $\nu_{U }$   observational frequencies of the millisecond pulsar XTE J1807-294. Here we study the $\nu_{L }$=$f(\nu_{U })$  correlation. The model function, however, is known only in parametric form. It is defined by the system:

\begin{eqnarray}
    \nu_{ L}(a,M,x)=\nu_{L}^{ obs}\\
\nu_{ U}(a,M,x)=\nu_{U}^{ obs}
\end{eqnarray}

where  $\nu_{L}^{ obs}$ and  $\nu_{U}^{ obs}$ are the observed values of the frequencies and  $x_{L}^{ obs}$ and $x_{U}^{ obs}$ are the radii of the orbits on which they originate. For the set of models considered here it is assumed that $x_{L}^{ obs}$=$ x_{U}^{ obs}$ , i.e. the simultaneous kHz QPOs originate on the same radius.

For given $M$ and $a$ , from eq. (1b) we can express  $x$ as function of $\nu_{U}^{ obs}$, $x=f(a, M, \nu_{U}^{ obs})$. This value is plugged in the model function for the lower frequency,  $\nu_{L }$ . This recipe allows us to predict the value of the lower frequency $\nu_{L }$ for each of the observed values of the upper frequency $\nu_{U}^{ obs}$. Thus predicted values are then compared to the observed ones.

\subsection*{3.2 Explicit formulae for the frequencies }
The application of the recipe described above requires explicit expressions for the frequencies   $\nu_{L }$ and  $\nu_{U }$ . As already mentioned, for the set of geodesic models studied here  $\nu_{L }$ and  $\nu_{U }$ are simple linear combinations (See Table 1) of the fundamental frequencies of motion of a test particle moving in the innermost parts of the accretion disk: the orbital (Keplerian) frequency $\nu_{\phi }$, the vertical $\nu_{\theta}$ and the radial $\nu_{r }$ epicyclic frequencies. Explicit formulas for the fundamental frequencies are given in the Appendix. They are obtained with Kerr metric.

The applicability of this metric for the description of the spacetime in the immediate vicinity of a neutron star is questionable and is one of the caveats of the current study. It is justified for the evaluation of the fundamental frequencies of neutron stars with specific angular momenta   lower than $0.4$ and masses close to or higher than  $2$ solar masses, ((T\"{o}r\"{o}k et al., 2010; Stuchl\'{\i}k \& Kolo\v{s}, 2015) and references therein).  (See also the discussion made on this issue by Stefanov, (2016).) The Kerr metric is rather attractive due to the simplicity of the analytic expressions  and the fact that it has only two free parameters,  mass and angular momentum. This is why we choose to work with it. Here we have worked with a wider range of values of the specific angular momentum, $a\in(0,1)$ and check whether the estimates that we obtain for it satisfy the constraint  $a<0.4$ when the uncertainties are taken into account at the end. The application of the Kerr metric is also legitimate for objects with $M>3M{\odot}$ but since this is the theoretical upper bound on the neutron star mass so heavy objects cannot be regarded as neutron stars.

\subsection*{3.3 $\chi^{2}$ variable }
The goodness of the fit is evaluated by the  $\chi^{2}$ test. In order to apply it we define the $\chi^{2}$ variable:

 \begin{eqnarray}
\chi_{L}^2(a,M)=\sum_i\left({\nu_L(a,M,\nu_{U}) -\nu_{L,i}^{obs},}\over \sigma_{L,i}\right)^2
\end{eqnarray}

where $\sigma_{L,i }$ is the uncertainty of the  -th observed value of the lower frequency. The optimal values of the free parameters $M_{ opt}$  and   $a_{ opt}$ are obtained through minimization of the function defined by eq. (2).

 \subsection*{3.4 Acceptability criteria and uncertainties }
 For N=7 pairs and 2 free parameters, i.e. 5 degrees of freedom,  $\chi^{2}$ should be in the range  $0\leq\chi^{2}_{min}\leq 11,1$ corresponding to $90$ percent confidence level. We compare the values of  $M_{ opt}$, $a_{ opt}$ and the minimum value of the chi-square variable  $\chi^{2}_{min}$ obtained by the different models in attempt of choosing the best one of them.

   \section*{4. Results and disscusion}
We have applied the $\chi^{2}$ test to 7 groups of data, from B to H (see table 1), thus including only data for which the kHz QPOs appear to be simultaneous.
The graphs that depict how the lower frequency  $\nu_{L }$ depends on the upper frequency  $\nu_{U }$ is given on Figure 1 for RP, RP1 and RP2 models, on Figure 2 for TP and TP1 models and on Figure 3 for TD and WD models. Table 3 contains the estimates for the optimal mass $M_{ opt}$ and the angular momentum $a_{ opt}$ according to the seven models that have been used.

In order to choose the best model explaining the observational frequencies of the kHz QPOs three different criteria that have to be met - $\chi^{2}_{min}$ that is low enough, $a_{ opt}$ that corresponds to the metric chosen and $M_{ opt}$ that is near the appropriate value for a neutron star.

As we mentioned above, within 90 percent confidence level  $\chi^{2}_{min}$ must be smaller than 11.1. According to this requirement the total precession models TP and TP1 as well as the tidal disruption model TD have to be discarded with  $\chi^{2}_{min}$=$12.50$, $12.36$ and $13.56$ correspondingly. These models give mass  $M_{ opt}$ that is more than acceptable especially when the standard errors are taken into account  -- between $2$ and $4M{\odot}$. 

The Kerr metric  can be applied for the description of the QPOs of slowly rotating neutron stars with $a<0.4$ (T\"{o}r\"{o}k et al., 2010; Stuchl\'{\i}k \& Kolo\v{s}, 2015). As it follows from this criterion, we are unlikely to explain successfully the kHz QPOs of the studied object using any of the relativistic precession models. Even though their $\chi^{2}_{min}$ show good agreement between the frequencies calculated by us using equations (1) and the data sets received by  Linares et al. (2005) - $\chi^{2}_{min}<11.1$ , the angular momenta received are near the upper bound for  this parameter, $ 0.9-1.0$. This conclusion does not change significantly, even if the standard error of the results is considered - still $a>0.4$.

  \begin{table}[htb]
  \begin{center}
  \caption{Estimates for the optimal mass $M_{ opt}$ and the angular momentum $a_{ opt}$ according to the applied models }
\begin{tabular}{ lc lc lc }
model                                                        & $\chi^{2}_{min}$                              & $a_{ opt}\pm\delta a$   & $M_{ opt}\pm\delta M$   \\
\hline \\
RP                                                             & $  1.50 $                                              & $1.0\pm 0.1$                 & $11\pm 3$      \\
RP1                                                           &  $ 9.07 $                                              & $0.9\pm 0.3$                   & $6\pm 2$\\
RP2                                                           & $  1.09 $                                              & $0.9\pm 0.1$                 & $13\pm 5$\\
 TP                                                            & $12.50$                                               & $0.0\pm 0.5$                   & $3\pm 1$\\
TP1                                                           & $12.36$                                               & $0.2\pm 0.5$                   & $4\pm 2$ \\
WD                                                            & $ 2.02 $                                              & $0.0\pm 1.4$                   & $3\pm 3$\\
  TD                                                           & $13.56$                                              & $0.0\pm 2.3$                   & $2\pm 2$\\
                     \end{tabular}
  \label{estimates}
  \end{center}
\end{table}
 The values of the mass that we obtain vary from $2M{\odot}$ to $13M{\odot}$ and the chosen by us Kerr metric is legitimate for object with such masses. The problem is that stars with masses greater than $3M{\odot}$ in the final stages of their evolution are considered black holes rather than neutron stars. Following this line we have to mention that relativistic precession models yield the biggest mass estimates – from $6\pm2M{\odot}$ for RP1 to $13\pm5 M{\odot}$ for RP2 i.e. they tend to overestimate the mass value and this trend is the worst for RP2. The total precession models TP and TP1 show mass that is appropriate for a neutron star, low angular momentum $0.0\pm0.5$ and $0.2\pm0.5$ respectively but bad $\chi^{2}_{min}>12.0$.

 For three of the models – TP, WD and TD the estimated value of the spin is near to zero, i.e. the optimal mass does not depend significantly on the angular momentum (on the rotation of the object). From them we discarded the tidal disruption model due to the big $\chi^{2}_{min}=13.56$, even if the mass calculated is the smallest one $ 2\pm2M{\odot}$. So, the only model we cannot entirely reject under the conditions of our investigation is the warped disk model, WD with$\chi^{2}_{min}=2.02$, $a_{ opt}= 0.0\pm2.3$ and $M_{ opt}=2\pm2M{\odot}$.

\begin{figure}[!htb]
  \begin{center}
    \centering{\epsfig{file=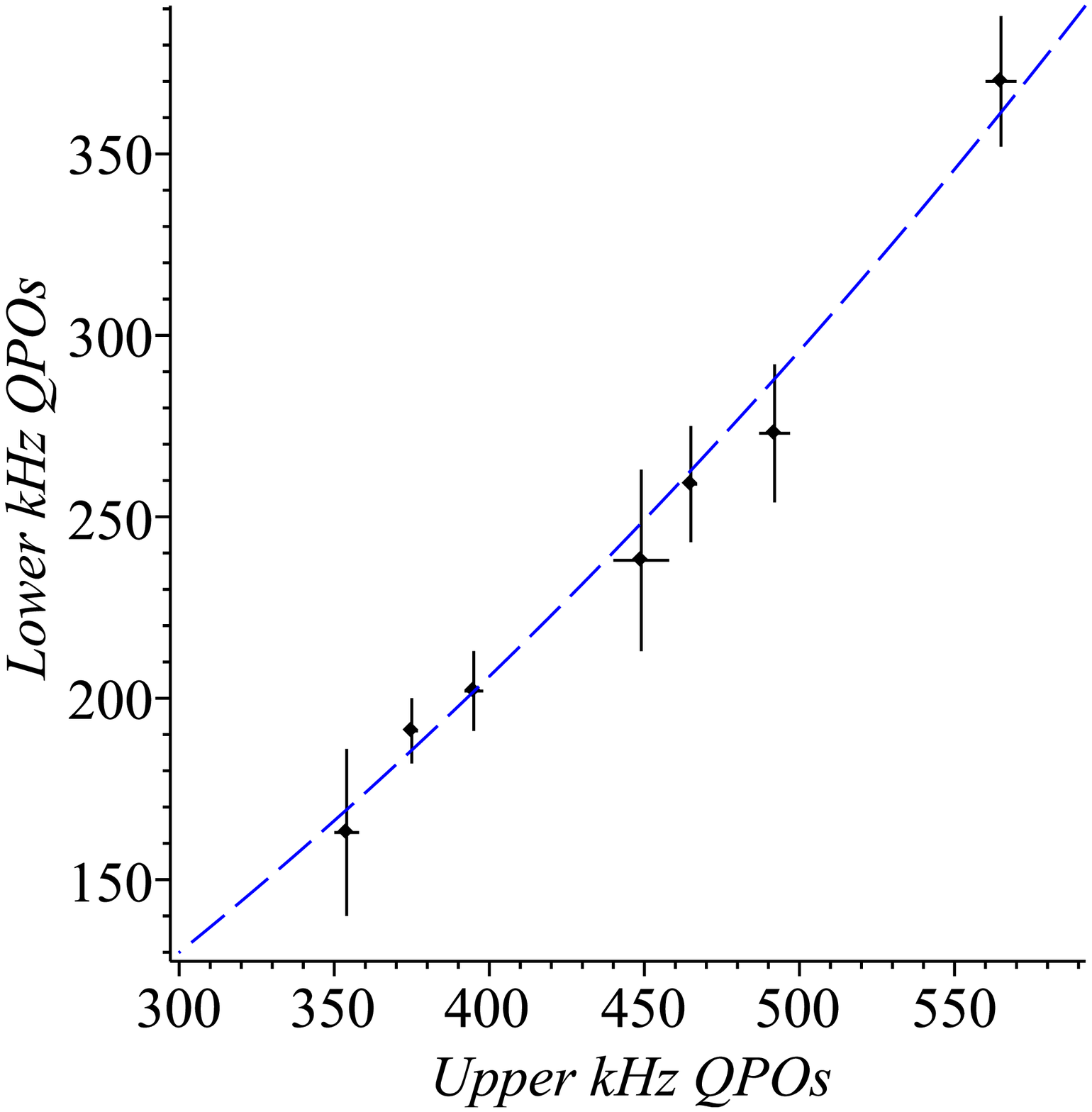, width=0.5\textwidth}\epsfig{file=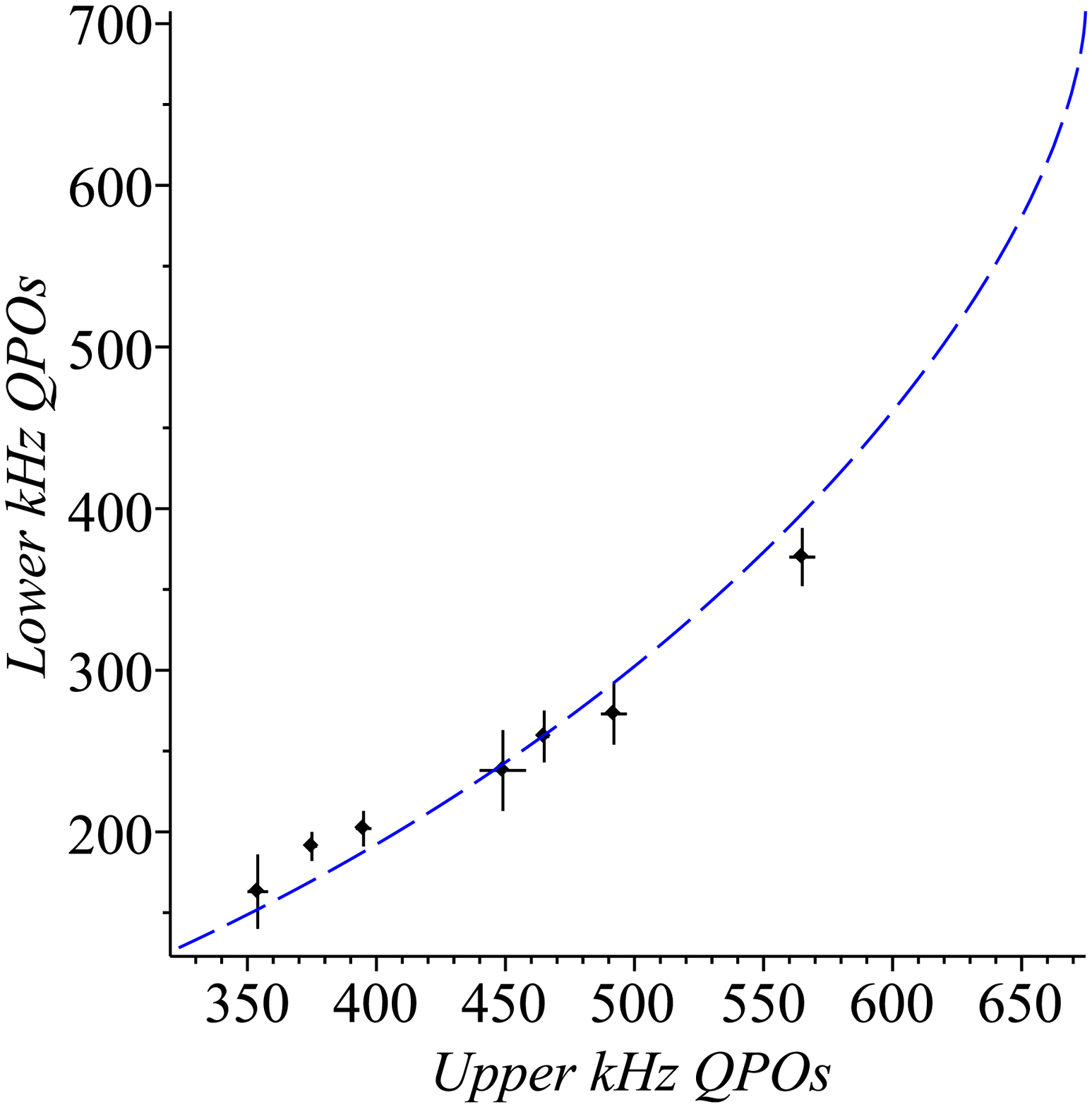, width=0.5\textwidth}\\ \epsfig{file=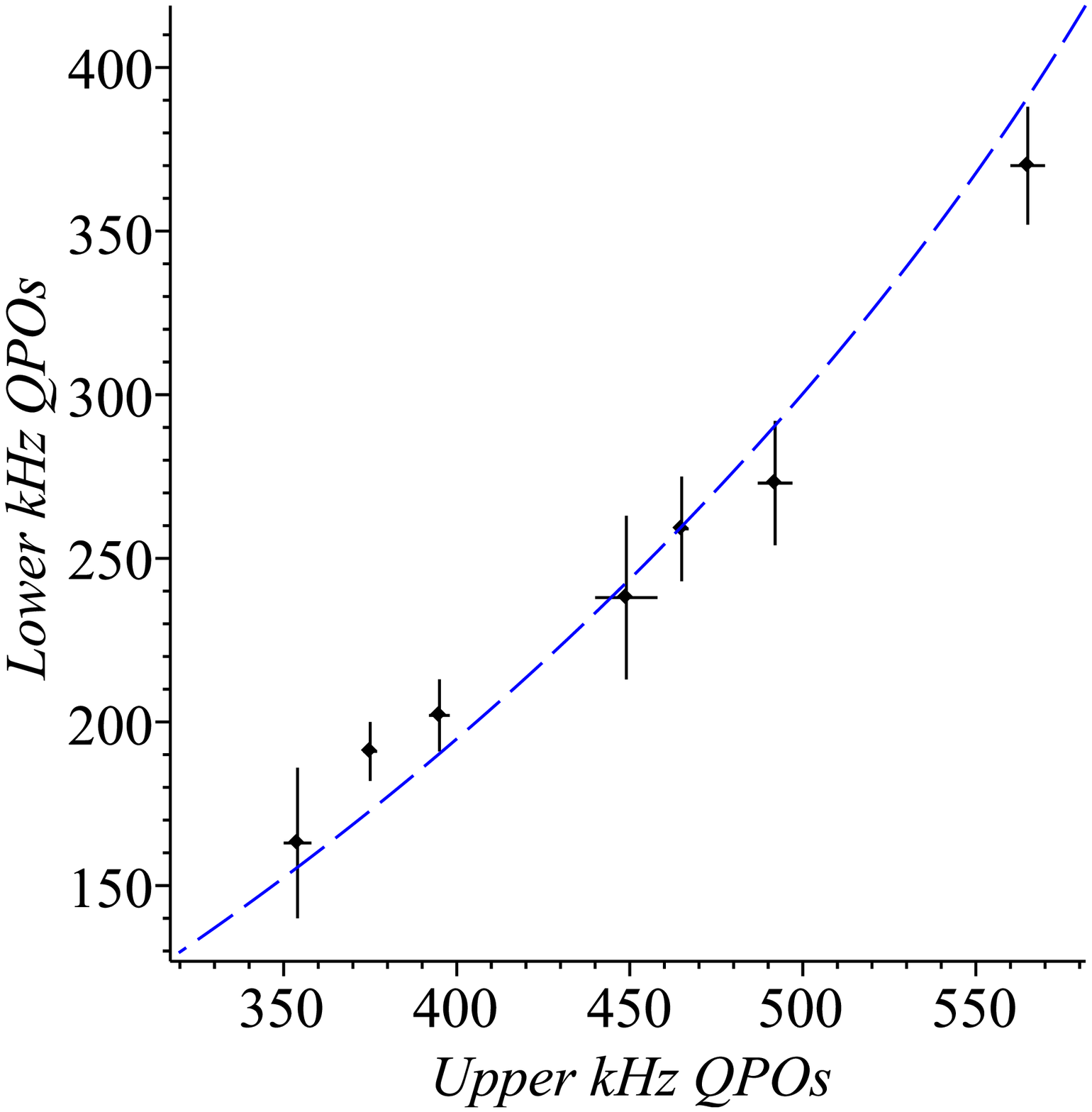, width=0.5\textwidth}}
    \caption[]{The $\nu_{L }(\nu_{U })$ dependence  according to the relativistic precession models. The  dashed  line represents the model function. The experimental points coming from the different groups - from B to H are given with their uncertainties.(a) RP model; (b) RP1 model; (c) RP2 model;}
    \label{countryshape}
  \end{center}
\end{figure}

\begin{figure}[!htb]
  \begin{center}
    \centering{\epsfig{file=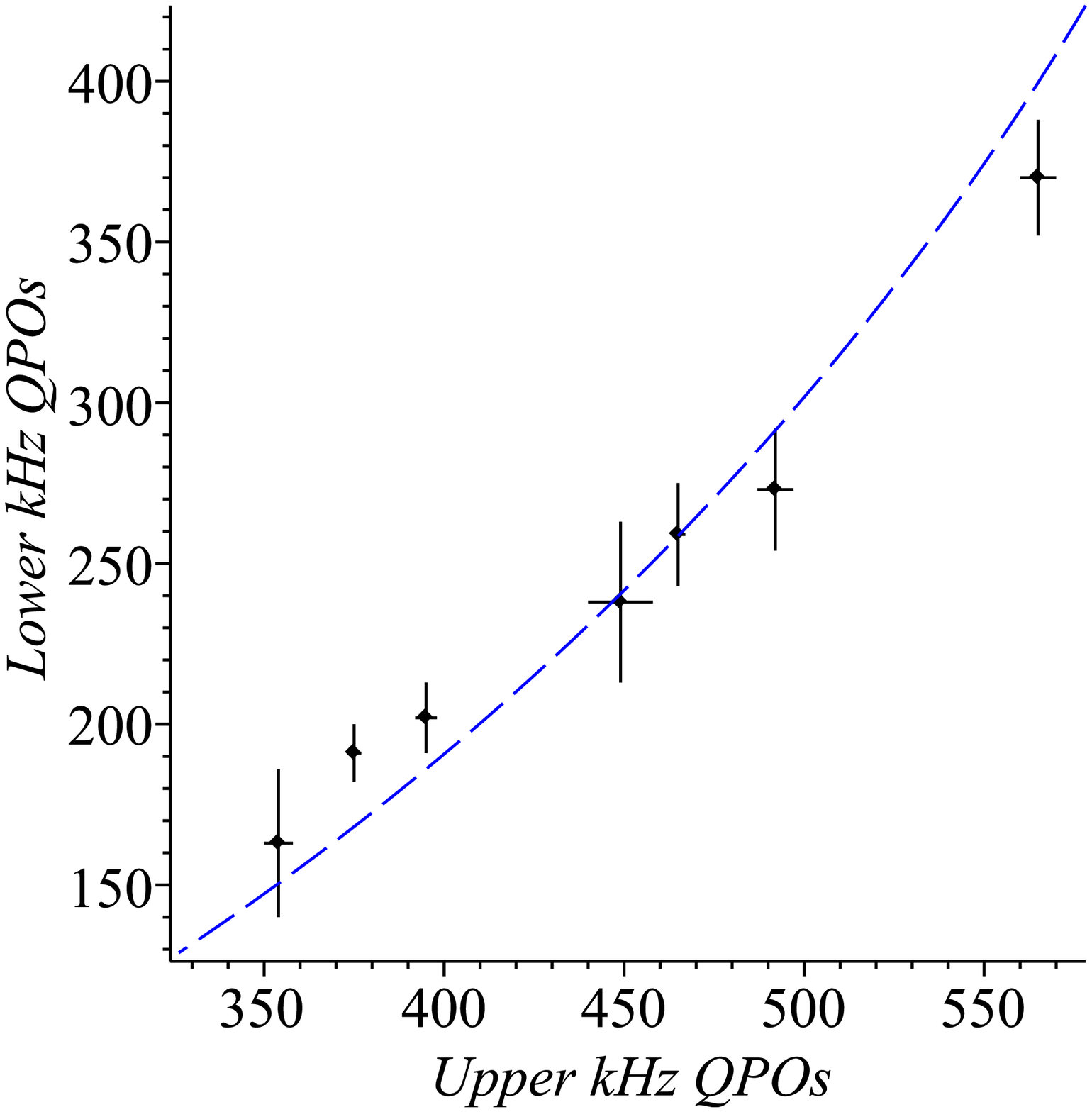, width=0.5\textwidth}\epsfig{file=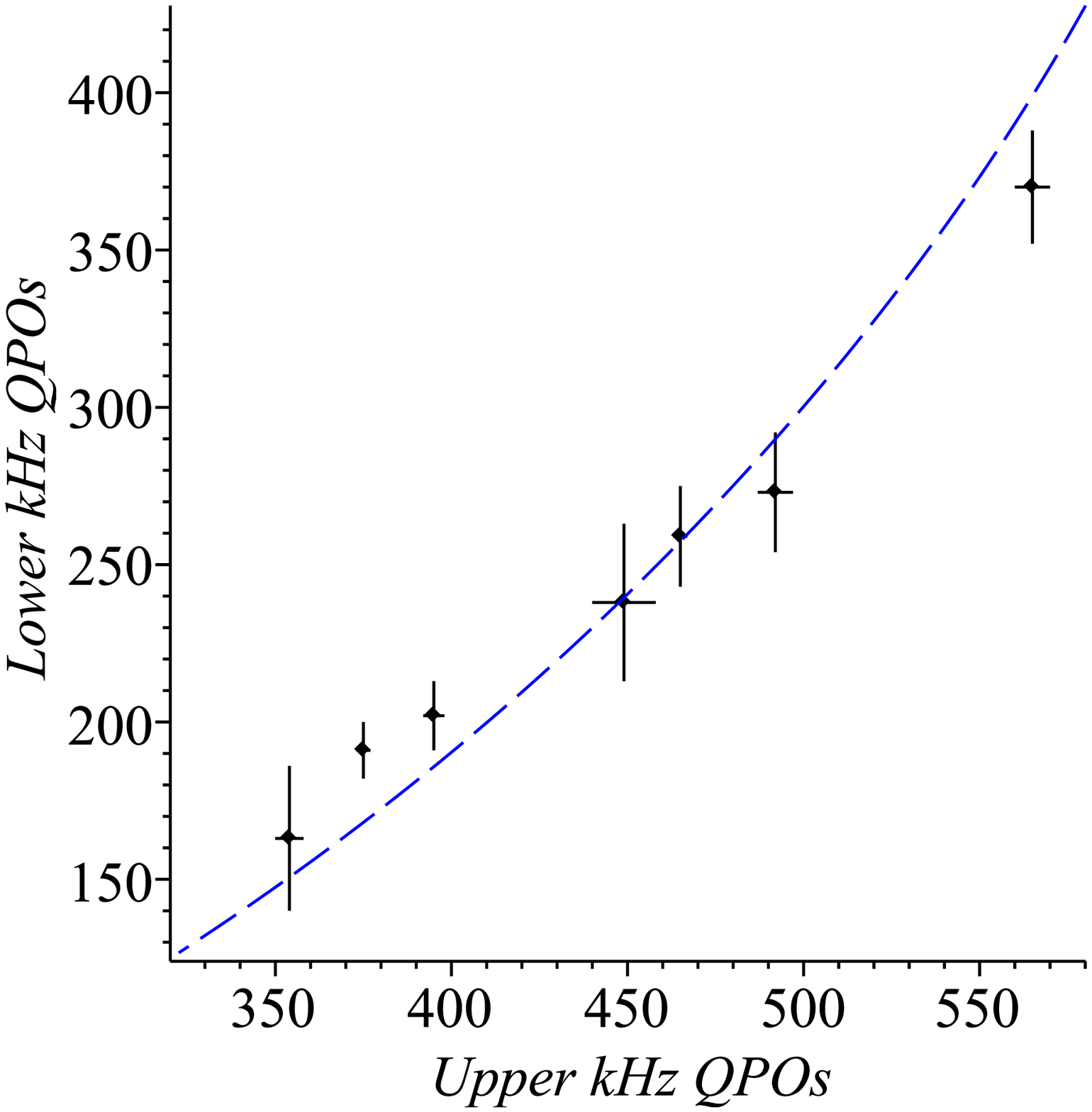, width=0.5\textwidth}}
    \caption[]{The $\nu_{L }(\nu_{U })$ dependence  according to the total precession models. The  dashed  line represents the model function. The experimental points coming from the different groups - from B to H are given with their uncertainties. (a) TP model; (b) TP1 model.}
    \label{countryshape}
  \end{center}
\end{figure}

\begin{figure}[!htb]
  \begin{center}
    \centering{\epsfig{file=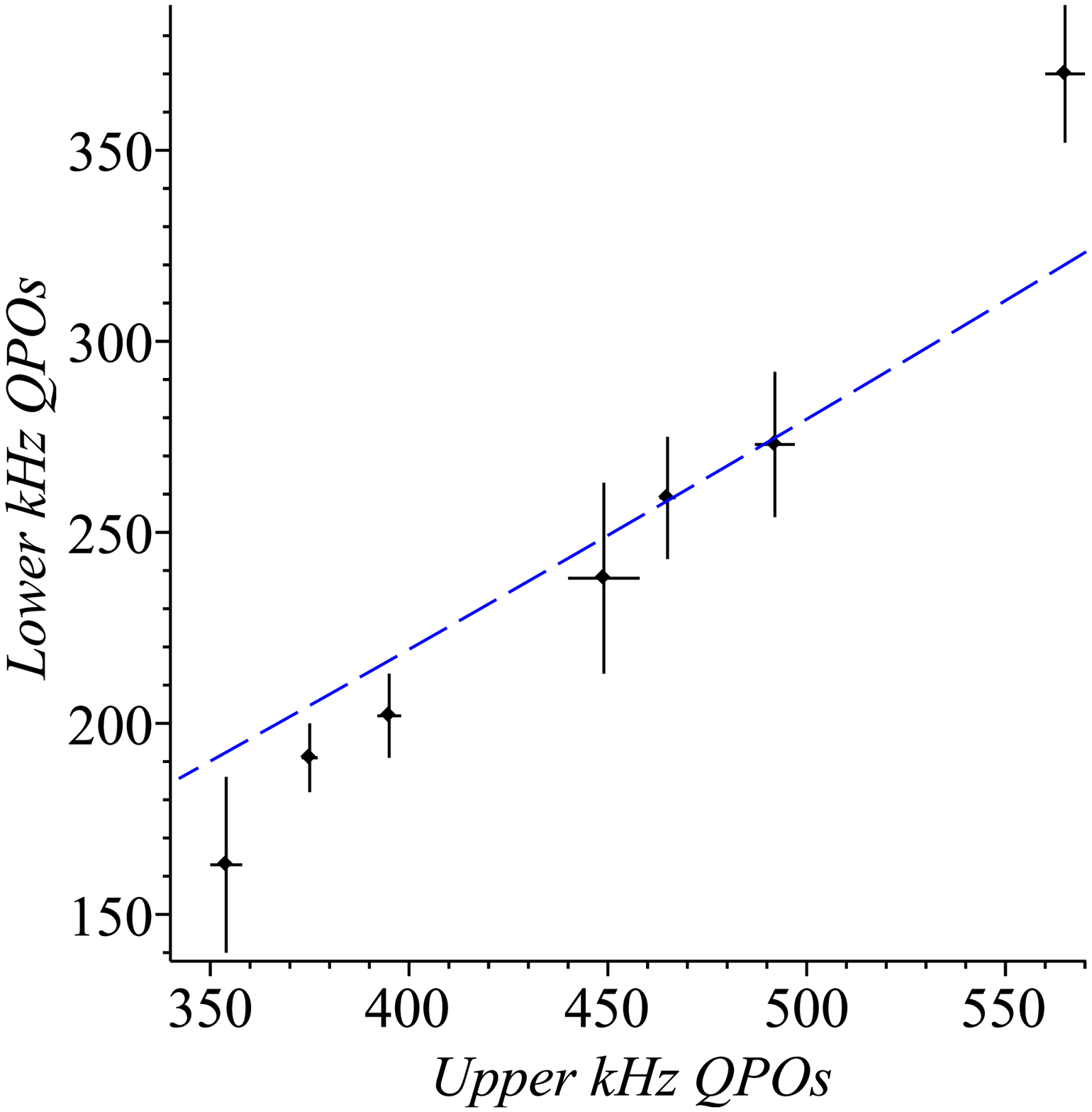, width=0.5\textwidth}\epsfig{file=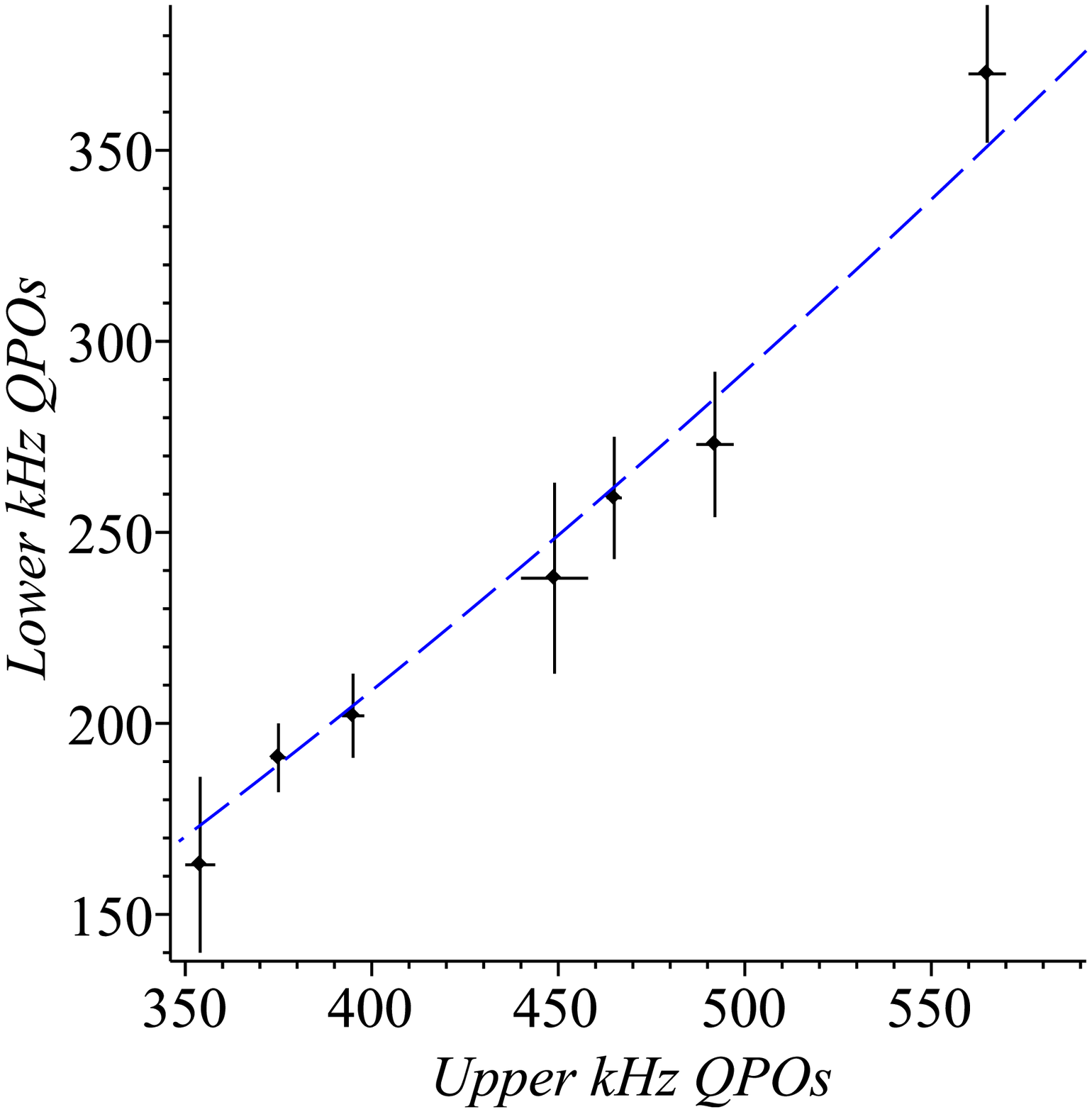, width=0.5\textwidth}}
    \caption[]{The $\nu_{L }(\nu_{U })$ dependence  according to the TD and WD models. The  dashed  line represents the model function. The experimental points coming from the different groups - from B to H are given with their uncertainties. (a) TD; (b) WD}
    \label{countryshape}
  \end{center}
\end{figure}

As we showed in (Tasheva \& Stefanov, 2018) for the case of the RP model the agreement between theory and experiment is better when the A data set of (Linares et al., 2005) is not taken into account. The RP model shows the best fitting of the data available according to the $\chi^{2}$ test (see also figure 1 (a)), but the WD model is the most plausible.
Our calculations suggest that having in mind the standard error according to WD model the mass of XTE J1807-294 is $2\pm2M{\odot}$. During the processing of the data through the $\chi^{2}$ test the uncertainties of the independent variable had to be neglected. If they are included, the results may change.

\section*{5. Conclusion}
We applied seven geodesical models to the microquasar XTE J1807-294. The dependence between the lower $\nu_{L }$ and the upper $\nu_{U }$ kHz twin QPOs found in the datasets of Linares et al. (2005)  tested by $\chi^{2}$ goodness-of-fit test allowed us to obtain estimates for the mass and the spin of the studied object - the ones for which $\chi^{2}=\chi^{2}_{min}$. The best $\chi^{2}_{min}=1.5$  received for the RP model unfortunately does not make it the best model due to high optimal mass $M_{ opt}=(11\pm3)M{\odot}$ and angular momentum $a_{ opt}=1.0\pm0.1$ that is not compatible  with the Kerr metric.

The relativistic precession models RP, RP1 and RP2 generally overestimate the mass – from $6\pm2 M{\odot}$ for RP1 to $13\pm5 M{\odot}$ for RP2 makig it too big for a neutron star. They also exhibit high angular momentum – near its upper bound, which invalidates the application of the Kerr metric.

The total precession models TP and TP1 yield good angular momenta $0.0\pm0.5$ and $0.2\pm0.5$ respectively. The obtained mass is suitable for a neutron star -  $3\pm1 M{\odot}$ and $4\pm2 M{\odot}$, if the standard error is included. The $\chi^{2}$ test though shows a bad agreement between the observed and the calculated frequencies $\chi^{2}_{min}=12.50$ and  $\chi^{2}_{min}=12.360$, respectively, i.e. these models also have to be rejected as non-adequate for description of the observed kHz QPO frequencies.

The tidal disruption model TD shows the worst $\chi^{2}_{min}=13.56$, the best mass  $2\pm2 M{\odot}$ and spin near to zero. These controversial results may suggest that some amendments of its concept are needed to make it more adequate to the object properties.

The warped disk model, WD is the only model that cannot be rejected – with $\chi^{2}_{min}=2.02$, $a_{ opt}=0.0\pm2.3$ and $Mopt=(2\pm2)M{\odot}$ it is the most plausible model that can be used in order to explain the kHz QPOs  of XTE J1807-294.

The reasons why so many models fail to explain the kHz QPOs  of XTE J1807-294 could be multiple - neglected uncertainties of the independent variable, non-simultaneous kHz QPOs, inappropriate choice of model. For further investigations we intend to try another type of metric accompanied by cross-over models – for example,the switch resonant model.

\section*{Acknowledgements}
The research is partially supported by the the Bulgarian National Science Fund under Grant N 12/11 from 20 December 2017.


\end{document}